\documentstyle[twoside,fleqn,npb,epsfig]{article}
\def\eps{\epsilon}

\def\as{\alpha_S}

\def\ktil{\widetilde k}
\def\kper{k_{\perp}}
\def\ep{\epsilon}
\def\cm{{\cal M}}
\def\lra{\leftrightarrow}
\def\beq{\begin{equation}}
\def\eeq{\end{equation}}
\def\beeq{\begin{eqnarray}}
\def\eeeq{\end{eqnarray}}
\def\nn{\nonumber}

\newcommand{\AmS}{{\protect\the\textfont2
  A\kern-.1667em\lower.5ex\hbox{M}\kern-.125emS}}

\title{Steps towards NNLO QCD calculations: collinear factorization at ${\cal
    O}(\as^2)$
\thanks{Talk given at 7th International Workshop on Deep
Inelastic Scattering and QCD (DIS 99), Zeuthen, Germany, April 19-23, 1999}
}

\author{M. Grazzini\thanks{Work supported by the Swiss National Foundation}\address{Institute for Theoretical Physics,
    ETH-H\"onggerberg\\CH 8093 Z\"urich, Switzerland}%
        }
 
\begin{document}

\begin{abstract}

I consider the singular behaviour of tree-level QCD amplitudes when the
momenta of three partons become simultaneously parallel and I
discuss the universal factorization formula that controls the singularities
of the multiparton matrix elements in this collinear limit.

\end{abstract}

\maketitle

\section{Introduction}


The properties of QCD in the soft and collinear limits play an
important role in doing higher order calculations. The singularities arising
in these limits
prevents a straightforward 
integration of matrix elements to
obtain physical cross sections. However, as far as
NLO calculations are concerned, the universality property of these
singularities provided a method to solve these problems. The singular
behaviour at this order is well known and is given in terms of ${\cal O}(\as)$
factorization formulae for tree level \cite{mangano} and one loop \cite{bern}
amplitudes in the soft and collinear limits.

As a first step towards NNLO calculations it would be of great importance to extend
this knowledge at ${\cal O} (\as^2)$. At this order one has to consider two
loop corrections, one loop corrections with real emission and double real
emission.
The singularities in two loop corrections have been
discussed in Ref. \cite{two}. The properties of one-loop amplitudes in the
soft and collinear limits have been recently obtained at all order in $\ep$ in 
Ref. \cite{one}. The singular behaviour of tree-level amplitudes has been
studied in Ref. \cite{glover,letter}. Here I will
concentrate on the collinear behaviour.

\section{Notation and kinematics}
\label{kinem}

We consider a generic scattering process involving 
final-state
QCD partons
with momenta 
$p_1, p_2, \dots$ Non-QCD partons $(\gamma^*, Z^0, W^\pm, \dots)$, carrying
a total momentum $Q$, are always understood. The corresponding tree-level
matrix element is denoted by
\beq
\label{meldef}
\cm^{c_1,c_2,\dots;s_1,s_2,\dots}_{a_1,a_2,\dots}(p_1,p_2,\dots) \;\;,
\eeq
where $\{c_i\}$, $\{s_i\}$ and $\{a_i\}$ are
respectively colour, spin and flavour indices. The matrix element squared
summed over final-state colours and spins will be denoted by 
$| \cm_{a_1,a_2,\dots}(p_1,p_2,\dots) |^2$.
If the sum over the spin polarizations of 
the parton $a_1$ is not carried out, we define the following
`spin-polarization tensor'
\beeq
\label{melspindef}
{\cal T}_{a_1,\dots}^{s_1 s'_1}(p_1,\dots) &\equiv & 
\!\!\!\!\!\!\!\!\!\!\sum_{{\rm spins} \,\neq s_1,s'_1} \, \sum_{{\rm colours}}
\cm^{c_1,c_2,\dots;s_1,s_2,\dots}_{a_1,a_2,\dots}\nn\\
&\times &\left[ \cm^{c_1,c_2,\dots;s'_1,s_2,\dots}_{a_1,a_2,\dots}
\right]^\dagger
\;.
\eeeq

In the evaluation of the matrix
element, we use conventional dimensional regularization ($d=4 - 2\ep$ space-time
dimensions, two helicity states for massless quarks and 
$d-2$ helicity states for gluons).

The relevant collinear limit at ${\cal O}(\as)$ is approached when the momenta
of two partons, say $p_1$ and $p_2$, become parallel. This limit is 
defined by setting:
\beeq
\label{clim}
&&p_1^\mu = z p^\mu + k_\perp^\mu - \frac{k_\perp^2}{z} 
\frac{n^\mu}{2 p\cdot n} \nn\\
&&p_2^\mu =(1-z) p^\mu - k_\perp^\mu - \frac{k_\perp^2}{1-z} \frac{n^\mu}{2 p\cdot n}
\eeeq
and sending $k_\perp \to 0$.
In Eq.~(\ref{clim}) the light-like ($p^2=0$) vector $p^\mu$ denotes the
collinear direction, while $n^\mu$ is an auxiliary light-like vector, which 
is necessary to specify the transverse component $k_\perp$ ($k_\perp^2<0$)
($k_\perp \cdot p = k_\perp \cdot n = 0$) or, equivalently, how the collinear 
direction is approached.
In the small-$k_\perp$ limit (i.e.\ neglecting terms that are less singular 
than $1/k_\perp^2$), the square of the matrix element in Eq.~(\ref{meldef})
fulfils the following factorization formula
\vspace*{-.3cm}
\beeq
\label{cfac}
&&| \cm_{a_1,a_2,\dots}(p_1,p_2,\dots) |^2 \simeq \frac{2}{s_{12}} \;
4 \pi \mu^{2\ep} \as\nn\\ 
&&\;\;\;\;\;\;\times \;{\cal T}_{a,\dots}^{s s'}(p,\dots) \;
{\hat P}_{a_1 a_2}^{s s'}(z,\kper;\ep) \;\;,
\eeeq
where $\mu$ is the dimensional-regularization scale.
The spin-polarization tensor ${\cal T}_{a,\dots}^{s s'}(p,\dots)$
is obtained by replacing the partons $a_1$ and $a_2$ on the right-hand side
of Eq.~(\ref{melspindef}) with a single parton denoted by $a$.
This parton carries the quantum numbers of the
pair $a_1+a_2$ in the collinear limit.

The kernel ${\hat P}_{a_1 a_2}$ in Eq.~(\ref{cfac}) is the $d$-dimensional
Altarelli--Parisi (AP) splitting function \cite{AP}. 
It depends on the momentum
fraction $z$ but also on
the transverse momentum $\kper$ and on the helicity of the parton $a$ in the
matrix element $\cm_{a,\dots}^{c,\dots;s,\dots}(p,\dots)$.
More precisely, ${\hat P}_{a_1 a_2}$ is in general a matrix
acting on the spin indices $s,s'$ of the parton $a$ in the 
spin-polarization tensor ${\cal T}_{a,\dots}^{s s'}(p,\dots)$.
Because of these {\em spin correlations}, the spin-average square
of the matrix element $\cm_{a,\dots}^{c,\dots;s,\dots}(p,\dots)$
cannot be simply factorized on the right-hand side of Eq.~(\ref{cfac}).

\section{Collinear factorization at ${\cal O}(\as^2)$}
\label{splitt}

In the following we are interested in the collinear limit at
${\cal O}(\as^2)$.
At this order there are two different collinear limits to be considered
\cite{glover}.
The first is when {\em two pairs} of parton momenta become independently
parallel. In this case the factorization formula is obtained by a
simple iteration of Eq. (\ref{cfac}). In the second case 
three parton momenta can simultaneously become
parallel. Denoting these momenta by $p_1, p_2$ and $p_3$, their most general
parametrization is
\beq
\label{kin3}
p_i^\mu = x_i p^\mu +k_{\perp i}^\mu - \frac{k_{\perp i}^2}{x_i} 
\frac{n^\mu}{2p \cdot n} \;, \;\;\;\;\;i=1,2,3 \;,
\eeq
where the notations are the same as in Eq.~(\ref{clim}).
Note that no
constraint
is imposed
on the longitudinal and transverse variables $x_i$ and $k_{\perp i}$.

It can be shown \cite{letter} that in the triple-collinear limit
$(k_{\perp i}\to 0)$ the 
matrix element squared still fulfils a factorization formula analogous to
Eq.~(\ref{cfac}), namely
\beeq
\label{ccfac}
&&\!\!\!\!\!\!\!\!\!\!| \cm_{a_1,a_2,a_3,\dots}(p_1,p_2,p_3,\dots) |^2 \simeq
\frac{4}{s^2_{123}} (4 \pi \mu^{2\ep} \as)^2 \nn\\
&&\;\;\;\;\;\;\;\;\;\;\;\;\;\times{\cal T}_{a,\dots}^{s s'}(xp,\dots) \;
{\hat P}_{a_1 a_2 a_3}^{s s'}
\;.
\eeeq
As in Eq.~(\ref{cfac}), the spin-polarization tensor 
${\cal T}_{a,\dots}^{s s'}(xp,\dots)$ is obtained by replacing the partons 
$a_1$, $a_2$ and $a_3$ with a single parent parton, whose flavour $a$ is 
determined by flavour conservation in the 
splitting process $a \to a_1+a_2+a_3$. 

The three-parton splitting functions ${\hat P}_{a_1 a_2 a_3}$ generalize 
the AP splitting functions in Eq.~(\ref{cfac}). The spin 
correlations produced by the collinear splitting are taken into account
in a universal way, i.e. independently of the 
specific matrix element on the right-hand side of Eq.~(\ref{ccfac}).
Besides depending on the spin of the parent parton, the functions 
${\hat P}_{a_1 a_2 a_3}$ depend on the momenta $p_1, p_2, p_3$.
However, due to their invariance under longitudinal boosts along
the collinear direction, the splitting functions can depend in a non-trivial
way only on the sub-energy ratios $s_{ij}/s_{123}$ and on the following
longitudinal and transverse variables:
\beq
\vspace*{-.1cm}
z_i = \frac{x_i}{x}\;\;\;\;\;\;\;\;
{\ktil}_i^\mu = k_{\perp i}^\mu - \frac{x_i}{x}\;
\sum_{j=1}^3 k_{\perp j}^\mu \;\;,
\vspace*{-.1cm}
\eeq
where $x=\sum_{i=1}^3 x_i$

The method used to derive these results
exploits the fact that
interfering Feynman diagrams obtained by squaring the amplitude 
$\cm(p_1,\dots,p_m,\dots)$ are collinearly suppressed when computed
in a physical gauge. Thus, in the evaluation of the triple
collinear limit we can limit ourselves to consider the diagrams in
Fig. 1.
\begin{figure}[htb]
\centerline{\epsfxsize=52truemm\epsfbox{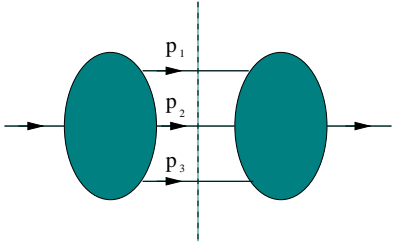}}
\vspace*{-.8cm}
\caption{Triple collinear limit: dominant diagrams in a physical gauge}
\vspace*{-.6cm}
\end{figure}
Details on the method and on our calculation are given in
Ref.~\cite{letter}.
The basic observation is that if we rescale the transverse momenta as $k_{\perp 
i}\to \lambda k_{\perp i}$ the matrix element squared
has the singular behaviour
\beq
| \cm_{a_1,a_2,a_3,\dots}(p_1,p_2,p_3,\dots) |^2 \sim 1/\lambda^4+...
\eeq
where dots stand for less singular contributions when $\lambda \to 0$.
To extract the singular behaviour one has to put on shell the parent parton leg in
Fig. 1.
Since $p_1+p_2+p_3=xp+{\cal O}(k_\perp)$, in order to do this we have to neglect
${\cal O}(k_\perp)$ terms from the amplitude which may affect in a non-universal way the singular
behaviour. But since we are interested in the {\em most singular} behaviour we 
can safely put $p_1+p_2+p_3\to xp$ and reconstruct the gauge invariant
spin polarization tensor in (\ref{ccfac}). Thus ${\hat P}_{a_1 a_2 a_3}$ can be
computed by evaluating the process independent diagrams in Fig.1 in the
collinear limit.
Our explicit results
are presented in \cite{letter}. We
find that when the parent parton is a fermion, spin-correlations are
absent. This is analogous to what happens at ${\cal O}(\as)$ and it is a
consequence of helicity conservation in the quark vector coupling.
On the contrary, in the case in which the parent parton is a gluon spin
correlations are highly non trivial.

A check of the calculation is provided by the strong-ordered limit. In this
limit the three partons become collinear sequentially and ${\hat P}_{a_1 a_2a_3}^{s s'}$ factorize in the product of two AP splitting functions.
A further non-trivial check of the calculation is provided by
supersymmetry. As for AP splitting functions, we find that ${\hat P}_{a_1 a_2 a_3}$ obey a
$N=1$ SUSY identity in the limit $\eps\to 0$ and when $C_F=C_A=2T_R$:
\beeq
&&\!\!\!\!\!\!\!\!\!\!\Big[{\hat P}_{{\bar q}_1q_2q_3}+(1\!\lra
\! 2)+(1\!\lra\! 3)\Big]\! +\!\Big[{\hat P}_{q_3g_1g_2}+(1\!\lra\! 3)\nn\\
&&\!\!\!\!\!\!\!\!+(2\!\lra\! 3)\Big]=P_{g_1g_2g_3}+\Big[{\hat P}_{g_3{\bar q}_1q_2}+5\mbox{ perm.}\Big]
\eeeq
which express the fact that the total quark (actually, gluino) and gluon decay probability are
the same.

The ${\cal O}(\as^2)$-collinear limit of tree-level QCD
amplitudes has been independently considered by Campbell and Glover \cite{glover}.
Taking for granted Eq. (\ref{ccfac}) they
neglected spin correlations and
computed only the spin-averaged splitting
functions by performing the limit of known amplitudes. We have compared our
results with those of Ref. \cite{glover} and found complete agreement.
\vspace*{-.1cm}
\section{Summary}
\label{summa}

I have discussed the three-parton collinear limit of tree-level QCD
amplitudes. In this limit the singular behaviour of the matrix element squared
is controlled
by process-independent splitting functions, which are analogous to the
Altarelli--Parisi splitting functions.

These splitting functions are one of the necessary ingredients to extend
QCD predictions at higher perturbative orders. In particular, they will be relevant
to set up
general methods to compute jet cross sections at NNLO. The knowledge of the
collinear splitting functions, when combined with a consistent analysis of
soft-gluon coherence properties, could also be used to improve the logarithmic
accuracy of parton showers available at present for Monte Carlo event
generators.

The results presented in this talk have been obtained in collaboration with
S. Catani.

\end{document}